\RequirePackage{fix-cm}
\documentclass[twocolumn,epjc3]{svjour3}  
\smartqed  
\RequirePackage{graphicx}
\journalname{Eur. Phys. J. C}
\usepackage[british]{babel}
\usepackage{hyphenat}
\usepackage{graphics}
\usepackage{multirow}
\usepackage{wasysym}
\usepackage{color}

\newcommand{\meg}{\ensuremath{\mu^+ \to e^+ \gamma}}

\RequirePackage{xspace}
\begin{document}

\title{\bf   The Quest for $\mu \to e \gamma$  and its Experimental Limiting Factors  at Future  High Intensity Muon Beams}

\author{G.~Cavoto\thanksref{inst1,inst2}, A.~Papa\thanksref{inst3}, F.~Renga\thanksref{e1,inst2},  E.~Ripiccini\thanksref{inst4} and C.~Voena\thanksref{inst2}} 

\thankstext{e1}{e-mail: francesco.renga@roma1.infn.it}

\institute{``Sapienza'' Universit\`a di Roma, Dipartimento di Fisica, P.le A.~Moro 2, 00185 Roma, Italy \label{inst1}
\and Istituto Nazionale di Fisica Nucleare, Sez. di Roma, P.le A.~Moro 2, 00185 Roma, Italy  \label{inst2}
\and Paul Scherrer Institut, 5232 Villigen, Switzerland \label{inst3}
\and Universit\'e de Gen\`eve, D\'epartement de physique nucl\'eaire et corpusculaire, 24 Quai Ernest-Ansermet, 1211 Gen\`eve, Switzerland \label{inst4}
}
\date{Received: date / Revised version: date}

\maketitle

\begin{abstract}
The search for the Lepton Flavor Violating decay \meg~will reach an unprecedented level of sensitivity within the 
next five years thanks to the MEG-II experiment. This experiment will take data at the Paul Scherrer Institut where 
continuous muon beams are delivered at a rate of about $10^8$ muons per second. On the same time scale, accelerator 
upgrades are expected in various facilities, making it feasible to have continuous beams with an intensity of $10^9$ 
or even  $10^{10}$ muons per second. We investigate the experimental limiting factors that will define the ultimate 
performances, and hence the sensitivity, in the search for $\meg$ with a continuous beam at these extremely high 
rates. We then consider some conceptual detector designs and evaluate the corresponding sensitivity as a function of 
the beam intensity.
\end{abstract}


\authorrunning{G.~Cavoto \emph{et al.}}
\titlerunning{The Quest for $\mu \to e \gamma$ at Future  High Intensity Muon Beams}
\flushbottom


\section{Introduction}
\label{sec:intro}

The search for lepton flavor violation in charged lepton decays like \meg~plays a crucial role in the search for
physics beyond the Standard Model (SM). The conservation of the lepton flavor is an accidental symmetry
in the SM and is generally broken in new physics (NP) models, which are already strongly constrained by the present limits.
The  discovery of neutrino oscillations already demonstrated that this symmetry is not exact, although the impact 
on the charged lepton sector is negligible, predicting  for the \meg~decay a branching ratio (BR)
of about $10^{-54}$, far away from the present experimental limit, $\mathrm{BR}(\meg) < 4.2 \times 10^{-13}$~\cite{meg_analysis}, 
obtained by the MEG collaboration at the Paul Scherrer Institut (PSI, Switzerland).

These features make the search for charged Lepton Flavor Violation (cLFV) very attractive: on one side, limits on 
$\mathrm{BR}(\meg)$ hugely impact the development of NP models; on the other side, an observation of this or any other cLFV decay would be 
an unambiguous evidence of NP~\cite{Lindner:2016bgg}, without any theoretical uncertainty.

In the search for cLFV in muon decays, a central role is played by the availability of high intensity continuous muon 
beams,\footnote{In this paper we concentrate on the search for cLFV in the decay of free muons, and in particular~\meg, 
which requires a continuous muon beam, and hence we do not discuss the efforts made to deliver very high intensity pulsed 
muon beams, e.g. for the search of $\mu \to e$ conversion in the Coulomb field of a nucleus.} and there are activities around 
the world~\cite{himb,music,pip-ii} to increase the beam rates to eventually reach $10^{10}$ muons per second.
In this context, it is crucial to understand which factors will limit the sensitivity of experiments to be run at these facilities in the 
future. In this paper we concentrate on the \meg~searches. After briefly reviewing the current experimental status and the 
ongoing efforts to build high intensity continuous muon beam lines, we will investigate
 the ultimate experimental resolutions and efficiencies which cannot be realistically surpassed with 
the current experimental concepts, even considering some incremental improvement in the detection techniques. 
Moreover, we will shortly discuss how these ultimate performances could be technically   reached. Finally, we determine 
the sensitivity which could be   obtained if the proposed strategies will be found to be technically feasible.

\section{Basics of $\mu \to e \gamma$ searches}
\label{sec:basics}

The largest step in $\mu \rightarrow e \gamma$ sensitivity was due to the transition  from  the search in cosmic muon  decays (rate of the order of Hz) to
muons from stopped pion beams  (four orders of magnitude higher rate) and  eventually to muon beams (two further orders of magnitude).
Within each beam configuration the improvements of the detector resolutions, which determine  the background rejection capability, were fundamental.

Muons are usually stopped in a target, in order to exploit the very clear signature of  a  decay at rest: 
an $e^+$ and a $\gamma$ in coincidence, moving collinearly back-to-back with their
energies equal to half of the muon mass ($m_{\mu}$/2 = 52.8 MeV). The searches are carried out by using positive muons: negative muons cannot be used, since they are captured by  nuclei  while  being stopped in the target.

There are two major sources of background events. One is the radiative muon decay (RMD),
$\mu^+ \rightarrow e \gamma \nu_e  \bar{\nu_{\mu}}$, when the positron and the photon are emitted almost back-to-back while the two neutrinos
carry off little energy. The other  is due to the  accidental coincidence of a positron from a Michel muon decay,  $\mu^+ \rightarrow e^+ \nu_e  \bar{\nu_{\mu}}$,
with a high energy photon, whose source might be either a RMD, the annihilation-in-flight  (AIF) of  a positron in a Michel  decay or the  bremsstrahlung from a positron.

To separate the signal from the various background events, four discriminating variables are commonly used. The positron energy $E_e$, the 
photon energy $E_\gamma$ and the relative angle $\Theta_{e\gamma}$ allow to reject both accidental and RMD events, while the further request of
a tight time coincidence between the positron and the photon (relative time $T_{e\gamma}$ = 0) helps by reducing the accidental background. It is
also important to notice that these variables are not correlated for accidental background events, and poorly correlated for RMDs on the scale of
the detector resolutions, while in  signal  events there is a precise expectation value for each of them. This makes it advantageous to use them separately 
in a statistical analysis, instead of combining them into an invariant mass.


In the four-dimensional space of these discriminating variables a signal region can be defined around their  expectation values for the signal events, with 
widths $\delta E_e$, $\delta E_\gamma$, $\delta T_{e\gamma}$ and $\delta \Theta_{e\gamma}$ which can be taken proportional to the corresponding resolutions. Hence, 
the impact of the resolution on each  variable can be quantified, considering the rate of accidental events falling in this signal region. According 
to~\cite{kuno-okada,bernstein}, this rate satisfies:
\begin{equation}
\label{eq:acc_rate}
B_{\mathrm{acc}} \propto \Gamma_\mu^2 \cdot \delta E_e \cdot (\delta E_\gamma)^2 \cdot \delta T_{e\gamma} \cdot (\delta \Theta_{e\gamma})^2
\end{equation}
 where $\Gamma_\mu$ is the muon stopping rate. This expression is derived considering the  photons from RMD, whose rate can be precisely predicted based on the RMD theoretical BR and the detector acceptance, with only minor corrections~\cite{meg_rmd}. For AIF photons, the absolute rate depends on the material crossed by the positrons along their trajectory, 
and hence on the details of the detector layout. 

A crucial element of Eq.~\ref{eq:acc_rate}  is the dependence on the square of $\Gamma_\mu$. Given the current detector resolutions, and with the large 
values of $\Gamma_\mu$ available at the present facilities, the accidental background is largely dominant over the prompt RMD contribution. Even imagining a sensible improvement
of the resolutions, this is likely to be the case also for the future facilities, when $\Gamma_\mu$ is increased by one or two orders of magnitude. Under these conditions, there are two regimes
for the expected experimental sensitivity. If one indicates with $B_{\mathrm{acc}} T$ the background yield in the signal region over the data-taking period of the experiment ($T$), the 
sensitivity improves linearly with the beam rate, as far as $B_{\mathrm{acc}} T\ll 1$ (efficiency-dominated regime). On the other hand, as soon as $B_{\mathrm{acc}} T \gg 1$, there is no advantage from a further increase of the $\Gamma_\mu$, since the ratio of the signal yield over the square root of the background yield remains constant (background-dominated regime). Indeed, the increased  pile-up of several muon decays  in the same  event would even deteriorate the detector 
performances. Hence, for a given detector, the optimal $\Gamma_\mu$ is the one for which no more than a few background events are expected over $T$. From another point 
of view, for a given $\Gamma_\mu$, the best compromise between resolutions and efficiency is the one giving a few expected background events, because it implies an optimal use of the available beam.


Some further considerations must be added  to the discussion above.

\begin{enumerate}
\item Tracking detectors can be used to determine precisely the positron direction, but photon detectors cannot provide by themselves a precise determination of the photon direction, 
to be used in the determination of the $\Theta_{e\gamma}$ angle. Hence, the following procedure is used: muons are stopped in a planar target, the intersection
of the positron track with the target plane (positron vertex) is taken as the muon decay point and the photon direction is taken as the vector going from the muon decay 
point to the photon detection point. Hence, the  $\Theta_{e\gamma}$ resolution is determined by the positron vertex resolution and the photon detection point resolution.
\item $B_{\mathrm{acc}}$ depends on the square of both the $E_\gamma$ and $\Theta_{e\gamma}$ resolution. In the first case this dependence arises from the quick drop of the RMD and 
AIF photon spectra at the kinematic end point. In the second case this can be understood by decomposing $\Theta_{e\gamma}$  in its  two independent projections, 
an azimuth angle $\phi_{e\gamma}$ and a polar angle $\theta_{e\gamma}$.
 This  dependence  implies that even a small improvement in the resolution of these  variables can have a significant impact on the sensitivity.
\item The rate of AIF photons, which tends to be dominant at the kinematical end point~\cite{kuno-okada}, depends on the material crossed by the positrons on their 
trajectories  (including positrons out of the detector acceptance and/or produced off-target). Hence, it is crucial to design the detector in order to have the lowest possible material
budget, not only in the tracking volume (as needed for  good positron resolutions), but in any region around the beam line, in particular near the target. The target
itself has to be considered as the main source of AIF photons.
\item Depending on the reconstruction techniques, further discriminating variables can be introduced to suppress the accidental background. If the photon detector allows
an even rough reconstruction of the photon trajectory, the likelihood of the photon and positron to come from the same vertex can be evaluated. It would help 
to further discriminate between signal and accidental background events. In this case, Eq.~\ref{eq:acc_rate} becomes~\cite{dejongh}:
\begin{equation}
\label{eq:acc_rate_conv}
B_{\mathrm{acc}} \propto \Gamma_\mu^2 \cdot \delta E_e \cdot (\delta E_\gamma)^2 \cdot \delta T_{e\gamma} \cdot (\delta \Theta_{e\gamma})^2 \cdot (\delta\Theta_\gamma)^2
\end{equation}
where $\delta\Theta_\gamma$ is the angular resolution of the photon detector.
\end{enumerate}


The last point above brings us to the discussion of the reconstruction techniques. Concerning the positron, the choice is between charged particle tracking in a magnetic spectrometer and  
calorimetry, and it is exclusively  driven by the achievable resolutions:  efficiencies are in fact comparable and potentially close to 100\% in both cases.
Conversely, for the photon the interplay between efficiency and resolution has to be carefully considered. A calorimetric technique was adopted in most of the past experiments, including 
MEG~\cite{meg-detector} with its Liquid Xenon (LXe) detector. This approach provides a large efficiency, only limited by the amount of material in front of the detector 
($\sim$~1~$X_0$ in MEG). However, a different technique was used in MEGA~\cite{mega}, a previous experiment performed at the Los Alamos National Laboratory: 
thin layers ($\sim$~0.1~$X_0$) of high-Z material were used to make the photon convert, and the resulting $e^+e^-$ pair was tracked in a magnetic field. The conversion 
efficiency is very low (few \%), but this technique provides a very precise energy measurement, an extremely precise conversion point measurement and some information 
about the direction of the photon.  Depending on the sensitivity regimes described above, these very good resolutions 
can compensate the loss in efficiency. 
 Therefore, if  $\Gamma_\mu$ is so low that the  $B_{\mathrm{acc}} $ can be reduced to a negligible level with a calorimetric technique, there is no real 
advantage from a large improvement of the resolutions, when it  comes at the price of a large efficiency loss. But in a high $\Gamma_\mu$ regime, when the calorimetric measurement 
would give too many background events, a strong improvement of the resolutions is the only way to really exploit the highest  $\Gamma_\mu$, because in this scenario it can compensate the concurrent efficiency loss.

\begin{figure*}
\centering
\resizebox{0.75\textwidth}{!}{
 \includegraphics{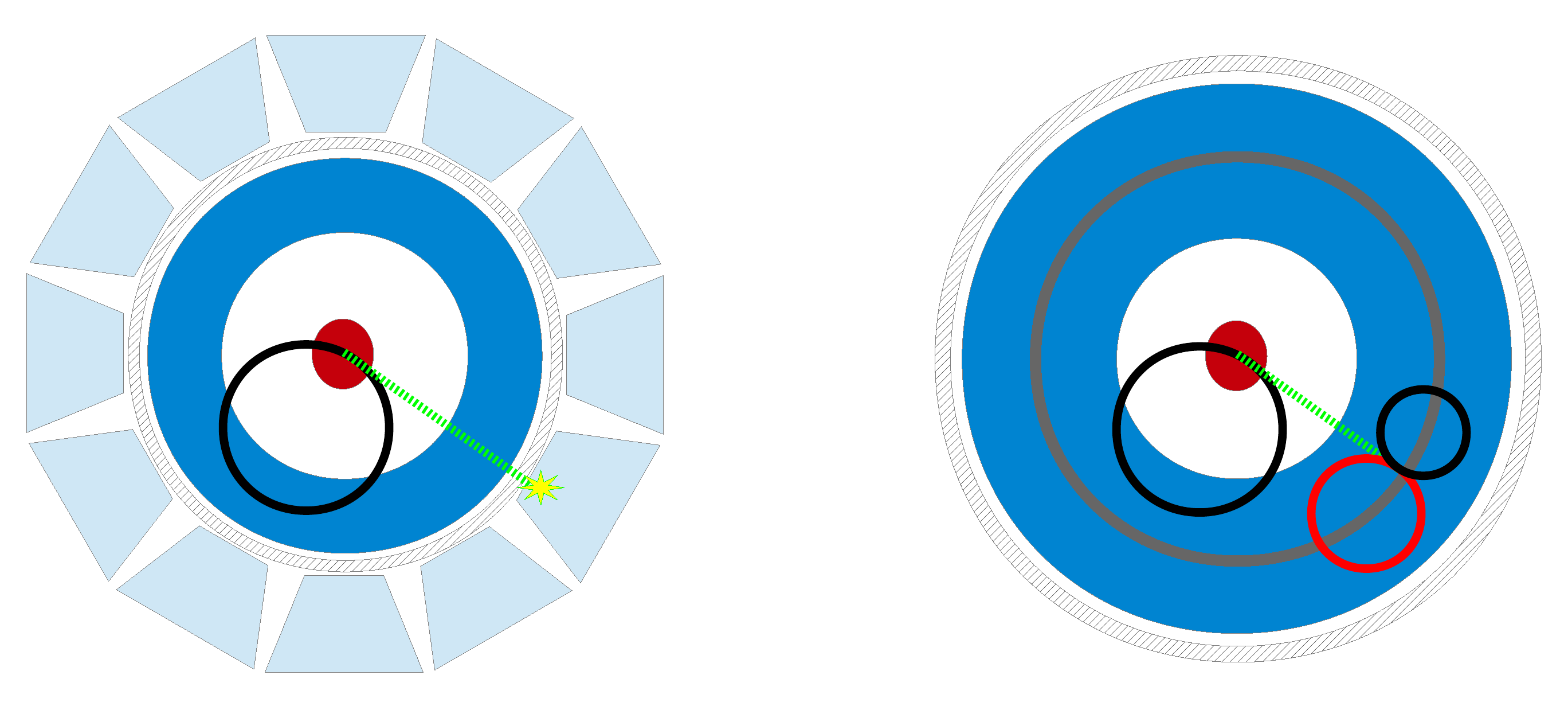}
}
\caption{Conceptual detector designs exploiting the calorimetric (left) or conversion (right) technique for the photon detection, and a tracking approach in a magnetic field 
for the positron reconstruction. Muons are stopped in a target (dark red ellipse) at the center of the magnet. Positron tracks from the muon decays (in black) are reconstructed in a 
tracking detector (dark blue), photons (in green) either produce a shower in a calorimeter (light blue) or are converted by a thin layer of high-Z material (in gray) into an electron-positron pair (in red and black, respectively) which is then reconstructed by an outer tracking detector. The magnet coil (hatched area) surrounds the tracking detectors. }
\label{fig:sketch_basics}
\end{figure*}

Fig.~\ref{fig:sketch_basics} shows how a typical detector can be designed to exploit either the calorimetric or conversion technique. 
In both cases, in order to measure precisely the time of the positron, fast detectors need to be placed at the end of the positron trajectory (not shown in the picture). 
If the photon is reconstructed in a calorimeter, a fast scintillator needs to be used in order to extract a good measurement of the photon time. Options for 
timing with the conversion technique will be discussed in detail in Sec.~\ref{sec:factors}.

A real-life example is the MEG experiment: positrons were tracked by a set of 16 planar drift chambers, in order to reconstruct their momentum and direction, 
and they finally reached a set of scintillating bars for timing purposes, while photons were detected inside a LXe calorimeter instrumented with PMTs, allowing to measure 
their energy, time and conversion position.

\begin{figure}
\centering
\resizebox{0.5\textwidth}{!}{
\includegraphics{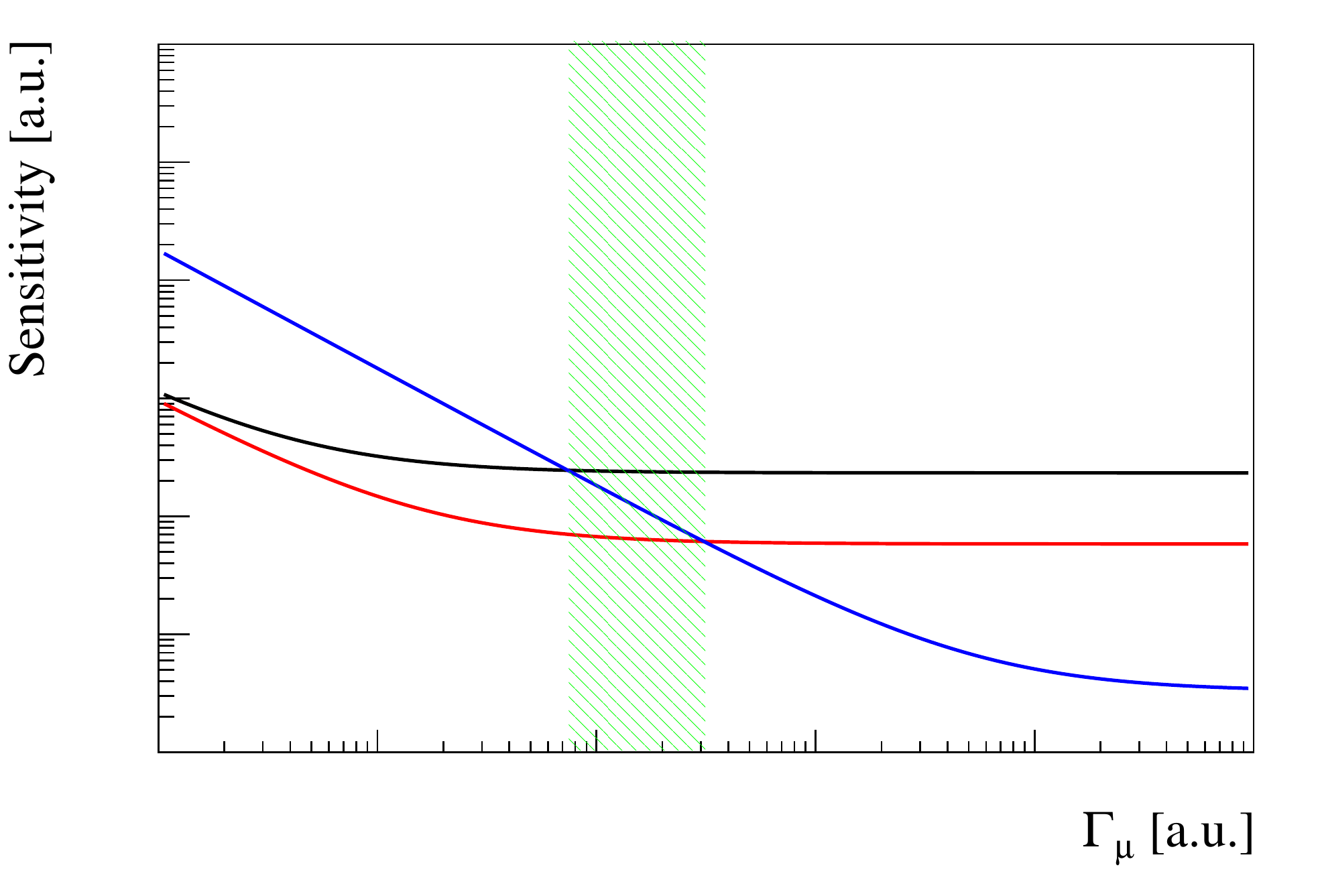}
}
\caption{Sensitivity trends as a function of the beam intensity, for a calorimetry-based design (black), a photon-conversion-based design with unchanged positron
resolutions (blue) and a calorimetry-based design with a factor two improvement in resolutions (red). See the text for a detailed description.}
\label{fig:sens_regimes}
\end{figure}

Some concepts presented above are illustrated in Fig.~\ref{fig:sens_regimes}. The  sensitivity is the smallest BR that can be excluded at some confidence 
level. The black line shows it for  a hypothetical experiment based on the calorimetric technique  against the $\Gamma_\mu$, for a fixed $T$. As discussed above, 
the sensitivity saturates at large $\Gamma_\mu$. The blue line shows instead the sensitivity of an experiment having 1/20 photon efficiency but 10 times better 
$E_\gamma$ and   $\Theta_{e\gamma}$ resolutions, like in a photon conversion approach. For low $\Gamma_\mu$ the calorimetric approach overcomes the photon conversion, 
but for very high $\Gamma_\mu$  the latter is advantageous. Notice that there is also an intermediate range (green-hatched area) where a moderate improvement in calorimetry 
(a factor 2 in resolutions for this example, red line) can bring this solution back to be preferable.  This is exactly what happened with the introduction of LXe calorimetry in MEG 
after the use of photon conversion in MEGA.

The MEG experiment is currently being upgraded (MEG-II, \cite{meg2}) with the same detector concept but several improvements which will push the sensitivity down of about one 
order of magnitude in three years of data taking. The main improvements with respect to MEG are: a 2~m-long  single-volume cylindrical drift chamber,  to improve the
tracking resolutions and the positron efficiency; a finer photon detector granularity at the inner face of the calorimeter, to improve the position and energy resolution and the
pile-up rejection capabilities of the detector; a highly segmented positron timing counter,  to improve the positron time resolution with multiple measurements
along the particle's trajectory.

\section{The next generation of high intensity muon beams}
\label{sec:beam}

The current best limit on the \meg~BR comes from the MEG experiment, operated at the $\pi$E5 beam line at PSI. Muons originate
from the decay of pions produced by a proton beam impinging on a graphite target. The $\pi$E5 channel is tuned to select positive muons with an 
average momentum of 28~MeV/c and a momentum bite of 5-7\% FWHM. This setup allows the selection of muons produced by pions decaying
right at the surface of the graphite target, providing high beam intensity and optimal rejection of other particles. A rate of
$10^8$ muons/s can be obtained, but is was limited to $3\times10^7$ muon/s in MEG, as this gave the best 
sensitivity, according to the discussion in Sec.~\ref{sec:basics}. In MEG-II  $\Gamma_\mu$ will be increased to $7 \times 10^7$ 
muons/s, thanks to the improved resolutions of the upgraded detectors. Another beam line  ($\mu$E4) is also operated at PSI, 
with the capability of delivering up to $5 \times 10^{8}$ muons/s.

In the meanwhile, an intense activity is ongoing at PSI and elsewhere to design channels for  continuous muon beams with
$\Gamma_\mu$ exceeding $10^{10}$ muons/s and possibly reaching $O(10^{11})$ muons/s. 

At PSI, the High-intensity Muon Beam (HiMB) project~\cite{himb} intends to exploit: 
\begin{enumerate}
\item an optimized muon production target;
\item a higher muon capture efficiency at the production target (26\% versus 6\% in the existing $\mu$E4 channel), thanks to a
new system of normal conducting capture solenoids;
\item a higher transmission efficiency (40\% versus 7\% in $\mu$E4), thanks to an improved design of the beam optics.
\end{enumerate}
Given the present $\Gamma_\mu$ in $\mu$E4, $5 \times 10^{8}$ muons/s in the experimental area, the goal of $O(10^{10})$ muons/s seems to be 
within reach.

At PSI, muons are produced on a relatively thin target (20~mm), since the beam has to be preserved for the subsequent spallation 
neutron source, SINQ. At RCNP in Osaka (Japan), the MuSIC project~\cite{music} makes use of a thicker target (200~mm), exploiting maximally
the much lower proton beam intensity. The target is surrounded by a high-strength solenoidal magnetic field in order to capture pions and 
muons with a large solid angle acceptance. Moreover, the field is reduced adiabatically from 3.5~T at the center of the target to 2~T 
at the exit of the capture solenoid in order to reduce the angular divergence of the beam and hence increase the acceptance of
the solenoidal muon transport beam line. Tests have been performed, showing that $\sim 10^6$ muons per Watt of beam power can be obtained. 
At the full beam power which is available at RCNP, a rate of $\sim 4 \times 10^8$ muons per second is expected at the production target, 
in the full momentum spectrum. The transport of the muons to the experimental areas and the selection of surface muons will reduce significantly this rate.
Nonetheless, it is a good example of the alternative approach for the production of intense continuous muon beams, with lower power of the
primary proton beam (400~W at MuSIC, to be compared with the 1400~kW power of the PSI proton accelerator) but a much higher muon yield
per unit of power, thanks to the thicker muon production target. A compromise between the two approaches could open interesting future 
perspectives for a further increase of the beam rates.

Ideas to perform searches for \meg~and $\mu^+ \to e^+ e^+ e^-$ have been also proposed in the framework of the PIP-II project
at FNAL~\cite{pip-ii,snowmass}. To the best of our knowledge, a realistic design of a continuous muon beam line at this facility and a reliable estimate of the 
achievable muon beam rates are not yet available in the literature. Nonetheless, there are indications that this facility could be competitive with the PSI
HiMB project.

These recent developments will give the possibility of running \meg~searches with muon beams one or two orders of magnitude more
intense than what is presently available.

\section{Experimental limiting factors}
\label{sec:factors}

In this Section we try to identify the factors which will ultimately limit the \meg~sensitivity of the next generation experiments. In this
respect, we will only marginally consider the intrinsic performances of the specific detection techniques (single hit resolutions, etc.), which
will be better discussed in the next sections. Conversely, our goals are to find the experimental factors (interaction with materials, etc.) 
which will not improve automatically with the technological evolution of the detectors, and to identify the experimental issues which will require 
a technological breakthrough in order to be addressed. 
Besides providing the basic information to estimate the potential sensitivity of the next generations of 
\meg~searches, this discussion will give some directions towards a more radical step forward.

\subsection{Efficiency}
\label{sec:eff}

The discrimination between signal and background events can be made through a maximum likelihood fit, as shown in MEG~\cite{meg_analysis},
with only a small loss of signal. The signal efficiency is therefore dominated by the positron and photon reconstruction efficiencies, $\epsilon_e$ and  $\epsilon_{\gamma}$.

The first element affecting $\epsilon_e$ and  $\epsilon_{\gamma}$ is the geometrical acceptance of the detector. Due to the back-to-back signature of the signal, the detector can be designed in such a  way that, for signal events, positrons never escape the detection if the photon is within the detector acceptance, or vice versa. So, one of the two sub-detectors 
unequivocally defines the detector acceptance. While in principle there is nothing preventing to have an almost full angular coverage, apart from a small region around 
the beam axis, costs can provide a strong limit. The MEG experiment, for instance, only had a 10\% acceptance, limited by the angular coverage 
of the (very expensive) LXe calorimeter. Though mitigated, this point could be relevant also for the innovative crystals we will discuss in Sec.~\ref{sec:calo}.

When  photons are  within the  acceptance, a fraction of them generate a shower before entering the detector.
This is mainly due to the material in front of the detector (photon detectors and the magnet coil of a positron spectrometer are typically placed 
in front of the active volume of the calorimeter). A reconstruction efficiency of $\sim 60\%$ was obtained in MEG, but different detector desings 
with lighter photon detectors could significantly improve this figure in the future.

Moreover, at larger $\Gamma_\mu$, the necessity of rejecting pile-up events implies some signal inefficiency. At $\Gamma_\mu$ $\sim 10^9$ muons per second, $B_{\mathrm{acc}}$ 
could be dominated by  the superposition of two RMD photons, with a total energy above 50 MeV, impinging on the photon detector, and the signal efficiency of
the necessary pile-up rejection algorithms could be relevant. An estimate of these effects largely depends on the specific detector design.

\begin{figure}
\resizebox{0.5\textwidth}{!}{
  \includegraphics{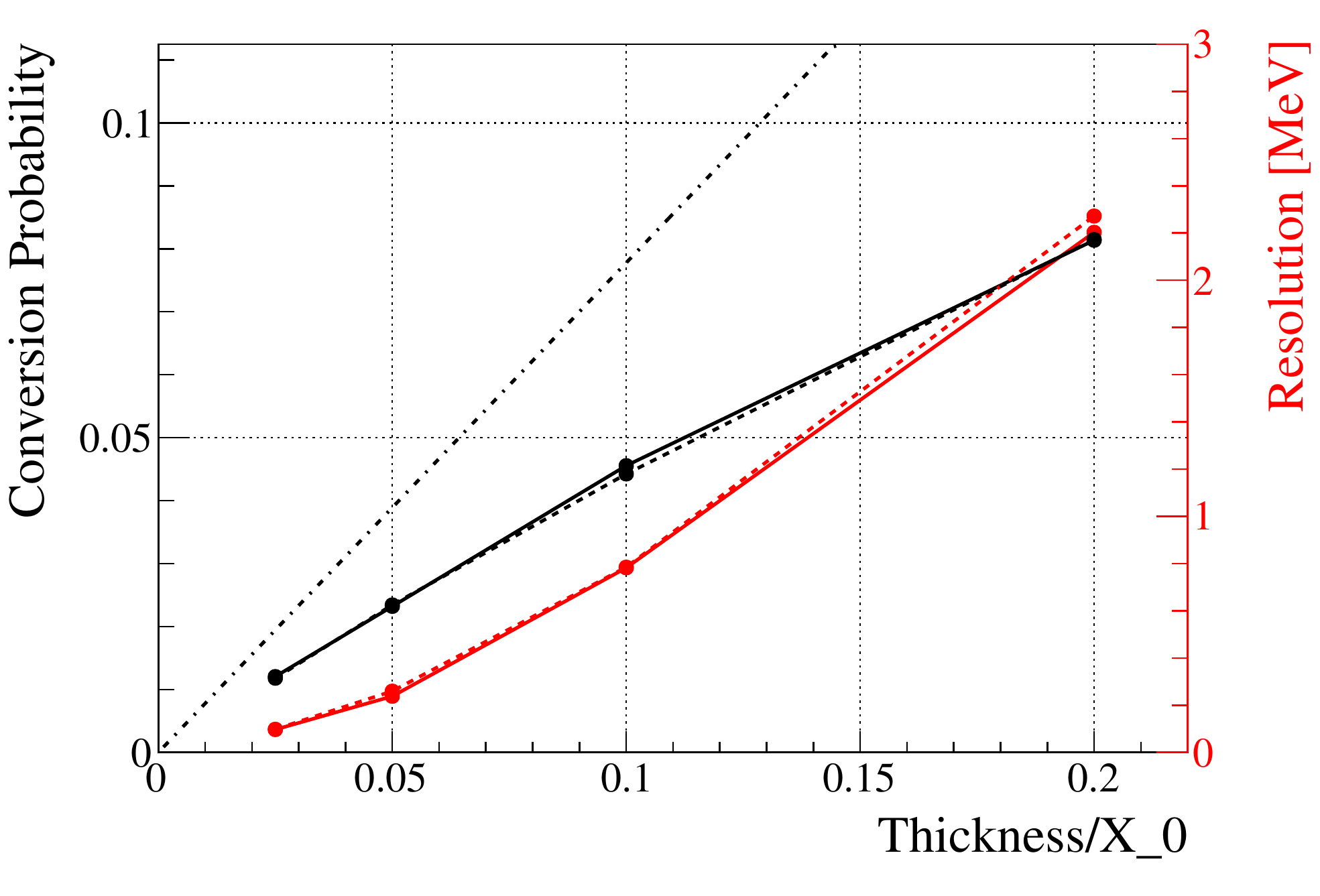}
}
\caption{The conversion efficiency (black, left axis) and the contribution to the energy resolution from the energy loss in the converter (red, right axis), for Lead (full lines) 
and Tungsten (dashed lines), as a function of the converter thickness (in units of radiation length). The dash-dotted line shows the asymptotic conversion probability, 
$7/9$ times the thickness in units of radiation length.}
\label{fig:Egamma_reso_simple}
\end{figure}

The situation is completely different if the photon conversion technique is adopted.  Thin converters are needed in order to preserve very good resolutions. It implies in turn 
a few percent  $\epsilon_{\gamma}$. In Fig.~\ref{fig:Egamma_reso_simple} the conversion probability for 52.8~MeV photons in lead and tungsten for different thicknesses are
shown. It must be noticed that, due to the relatively low energy of the photons, this probability is lower than the high-energy asymptotic value ($7/9$ times the thickness
in units of radiation lengths). Moreover, both the electron and the positron produced in the photon conversion have to be sufficiently energetic to be efficiently 
tracked. Considering that a typical tracking detector will have a few mm granularity along the track direction, only tracks with at least a few MeV can be reconstructed in a 
magnetic field of 1~T. Although the magnetic field can be optimized, a low momentum cutoff is unavoidable and, for instance, the requirement that the electron and the positron 
energies are both larger than 5~MeV further reduces $\epsilon_{\gamma}$  by  a factor $\sim 20\%$.

Concerning the positron from the muon decay, both tracking and calorimetry usually provide very large $\epsilon_{e}$. Inefficiencies, however,  can arise when the track is  propagated from the last measurement point in the tracking detector to the positron timing detector. Multiple Coulomb scattering (MS)  or energy loss  ($\Delta E$) might be dramatic and introduce large inefficiencies in 
the matching between the spectrometer and the timing detector. In MEG this effect was particularly important and reduced  $\epsilon_{e}$ by a factor of two.

\subsection{Photon energy}
\label{sec:Eg}

{\bf Calorimetry.} The $E_\gamma$  resolution is dominated by the photon statistics. Hence, the light yield determines the choice of the scintillator to be used, along with the fast 
response that is needed in order to reach a very good time resolution. Tab.~\ref{tab:scintillators} summarizes the relevant properties of some  state-of-the-art scintillating materials. 

 A degradation of the resolution due to the stability of the energy scale can be avoided with an accurate and frequent  multi-channel calibration. In MEG this enabled the
$E_\gamma$ scale to be kept stable to within 0.2\%.

\begin{table}[h!t]
\begin{center}
\caption{\label{tab:scintillators} Properties of state-of-the-art scintillators relevant for the application on \meg~searches.}
\begin{tabular}{lccc}
{\bf Scintillator} & {\bf Density]} & {\bf Light Yield} & {\bf Decay Time}\\
& {\bf [g/cm$^3$]} & {\bf [ph/keV]} & {\bf [ns]}\\
\hline  
\hline  
LaBr$_3$(Ce) & 5.08 & 63 & 16 \\
LYSO & 7.1 & 27 & 41 \\
YAP & 5.35 & 22 & 26 \\
LXe & 2.89 & 40 & 45 \\
NaI(Tl) & 3.67 & 38 & 250 \\
BGO & 7.13 & 9 & 300 \\
\hline
\end{tabular}
\end{center}
\end{table}

{\bf Pair conversion.} The limiting factor of the $E_\gamma$ resolution is the interaction of the $e^+e^-$ pair within the material of the photon converter itself.
Indeed, just after the conversion, the electron and the positron lose energy before exiting the converter. The  $\Delta E$ fluctuation predominantly contributes to the
resolution, since  $E_\gamma$  is estimated as the sum of the $e^+$ and $e^-$ energies (in some previous studies like~\cite{caltech} this contribution was disregarded~\cite{caltech-private}). According to our GEANT4~\cite{geant4} simulations, a 280~$\mu$m Pb layer ($\sim 5\%~X_0$), with photon conversions happening uniformly along the thickness of the converter, would give a resolution of $\sim 240$~keV 
in the limit of perfect tracking of the $e^+e^-$ pair. In Fig.~\ref{fig:Egamma_reso_simple} the contribution of the material effects to the resolution is also shown versus 
the layer thickness for lead  and tungsten, along with the total conversion probability. The resolution is evaluated as a truncated RMS of the reconstructed energy distribution, 
after discarding 20\% of the events in the low energy tail. 

Considering that a lower thickness improves the $E_\gamma$ resolution but also lowers  $\epsilon_\gamma$, an optimization is necessary. As pointed out in~\cite{dejongh}, 
the background rate is expected to scale with the third power of the converter thickness $t$, while  $\epsilon_\gamma$ scales linearly. 
So, one can try to maximize a Punzi figure of merit~\cite{punzi}:
\begin{equation}
f.o.m. = \frac{\epsilon_\gamma}{n_\sigma/2 + \sqrt{B_{\mathrm{acc}}}} \propto \frac{t}{n_\sigma/2 + \sqrt{B_0(t/t_0)^3}} 
\end{equation}
where $B_0$ is the background yield expected with $t = t_0$ at given $\Gamma_\mu$ and $T$. Typical choices of 
$n_\sigma$ are $2$  or  $3$. This function has a maximum for a $t$ such that the number of expected background events is $B_0(t/t_0)^3 = n_\sigma^2 = 4 \sim 9$. 
It indicates that, if allowed by the available beam intensity, the converter should be designed to yield  from a few to $\sim 10$ accidental background events. If the background yield is much higher, it is convenient to reduce  $t$, in order to improve the 
resolutions. If it is significantly lower, it is worth increasing $t$ to get a higher $\epsilon_\gamma$, to the detriment of the resolution. If it is much lower, a 
calorimetric approach is likely to perform better with that beam intensity.

The choice of the material of the converter has  to be considered too. The $\Delta E$ fluctuations are $\propto Z\rho/A$, while 
the resolution on the photon angle for vertexing is determined by the MS on the converter, which depends upon the square root of the number of radiation lengths 
($\sqrt{x/X_0} \propto \sqrt{Z^2\rho/A}$). From Eq.~\ref{eq:acc_rate_conv} we get $B_{acc} \propto Z^4\rho^3/A^3$. Given that the conversion efficiency is proportional
to the number of the radiation lengths ($\epsilon \propto Z^2\rho/A$), we conclude that in the background-dominated regime (where the Punzi f.o.m. can be 
approximated with $\epsilon/\sqrt{B_{\mathrm{acc}}}$) the sensitivity improves with increasing $\sqrt{\rho A}$, while it obviously goes with $Z^2\rho/A$ in the 
efficiency-dominated regime. Hence, dense, large-$Z$ materials are favored as converters, and Lead or Tungsten are typical choices.

\subsection{Positron energy}
\label{sec:Ee}

The  material in front of the  positron detector ultimately limits the $E_e$  and positron angular resolutions by MS and $\Delta E$ fluctuations.

The detector technology adopted for a tracking approach is therefore relevant: while gaseous detectors have been the choice for both MEGA and MEG, a silicon vertex tracker is used for the search of $\mu^+ \to e^+ e^+ e^-$ by the Mu3e  Collaboration~\cite{mu3e-tracker}, and a similar design has been suggested for future \meg~searches \cite{caltech}. State-of-the-art silicon pixel detectors can reach very good position resolutions ($\sim 3~\mu$m), with a thickness of 50~$\mu$m Si + 25~$\mu$m Kapton per layer~\cite{hvmaps}, corresponding to $\sim 10^{-3}$ radiation lengths per layer. On the other hand, the complete drift-chamber spectrometers of MEG or MEG-II amount to less than $3 \times 10^{-3}$ radiation lengths over  the whole track length within the tracking volume, nonetheless material effects gave a significant contribution in MEG and will almost be  dominant in MEG-II. It clearly indicates that more than a few silicon layers cannot be used: indeed, simulations~\cite{caltech} point toward $E_e$ resolutions of $\sim$~200~keV, which are not competitive 
with what can be obtained with gaseous detectors~\cite{meg2}.
We then  believe that the positron spectrometer of a next-generation \meg~experiment  has to incorporate an extended tracking region (dozens of cm of track length for a magnetic field of $\sim$~1~T) with dozens of measurement points and a few  $10^{-3}$ radiation lengths material budget, providing a  single  hit resolution of $\sim 100~\mu$m and hence a momentum resolution of $\sim 100$~keV ~\cite{meg2,dcproto}.

\subsection{Relative angle $\Theta_{e\gamma}$ }
\label{sec:Thetaeg}

The relative angle  $\Theta_{e\gamma}$ is measured  by combining the positron angle, the  photon conversion point and the positron vertex on the target.

The MS  and $\Delta E$  in the target and the material in front  of the spectrometer (e.g. the inner wall of a gas chamber) limit the measurement of the positron track direction. The target material  is also a relevant  source of  AIF photons pointing towards the photon detector. However, the target has to be thick enough to provide a good stopping power for muons. A good compromise has been obtained by slanting the target with respect to the beam axis  (in  MEG the target normal vector makes an angle $\alpha \sim 70^\circ$ with the beam axis, which will be increased to $76^\circ$ in MEG-II). In this configuration, the effective thickness seen by muons is magnified by a factor $1/\cos(\alpha) \sim 3$, while positrons emitted at the center of the detector acceptance ($90^\circ$ with respect to the beam axis) see a thickness magnified only by a factor $1/\sin(\alpha) \sim 1.06$. In Tab.~\ref{tab:straggling_ms_target} we show the angular uncertainties induced by targets of different materials.
GEANT4 simulations have been used to determine, for each material, the target thickness providing 90\% stopping power and the distribution of the stopping depth, used then in the simulation of the positron energy loss. In the best case, a contribution to the angular resolutions of about 3~mrad is found. It should be noticed that, due to the target geometry, this contribution depends on the angular acceptance of the detector. We assume here a full acceptance in $\phi_e$ and $\pm \pi/4$ acceptance in $\pi/2 - \theta_e$. Some strategies to reduce this contribution are discussed in Sec.~\ref{sec:target}.

\begin{figure}
\resizebox{0.5\textwidth}{!}{
  \includegraphics{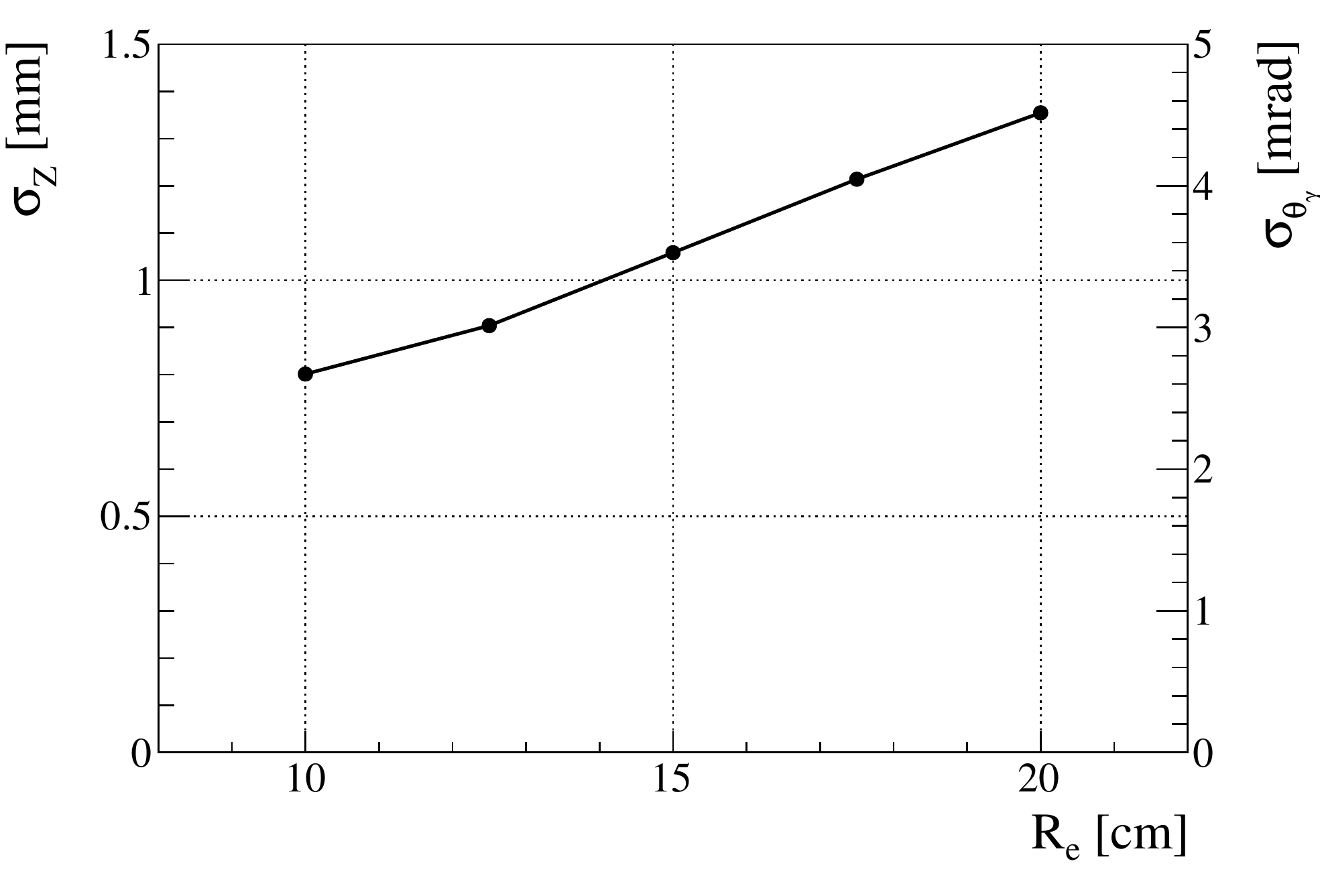}
}
\caption{Vertex resolution in the $Z$ coordinate as a function of the inner radius of the tracking detector. It gives a contribution to the $\theta_{e\gamma}$ resolution
which equals approximately $\sigma_{Z}/R_\gamma$, being $R_\gamma$ the radius of the photon detection point. The right axis shows this contribution for $R_\gamma = 30$~cm.
It has to be added to the contribution from the positron $\theta$ angle reconstruction.}
\label{fig:reso_proj}
\end{figure}

Due to the back-propagation of the track from the measured points  to the target, $E_e$  and  positron angular uncertainties at the inner layer of the spectrometer also translate into vertex position uncertainties at the target, which increase with the radius $R_e$ of the inner tracking layer. In this respect, it is crucial to have this first layer as close as possible to the target. These effects are illustrated in Tab.~\ref{tab:straggling_ms_gas}, where different scenarios are considered for $R_e = 20$~cm, and in Fig.~\ref{fig:reso_proj}, where the dependence on $R_e$ is shown, as an example, for the vertex resolution in the $Z$ coordinate. We assumed here the tracking resolutions expected for the MEG-II drift chamber, the MS due to Helium and a 25$~\mu$m Kapton foil just in front of the inner tracking layer, a magnetic field of 1~T, tracks emitted in the acceptance and target configuration of MEG-II and a photon detector placed at $R_\gamma = 30$~cm.

\begin{table*}[h!t]
\begin{center}
\caption{\label{tab:straggling_ms_target} Angle and energy uncertainties introduced by material effects in the target, for different target materials. A $76^\circ$ slant angle 
is assumed. The chosen thickness is the one providing 90\% muon stopping power.}
\begin{tabular}{lcccc}
\multirow{2}{*}{\bf Material} & \multirow{2}{*}{\bf Thickness [$\mu$m]} & \multicolumn{3}{c}{\bf Resolutions}\\  
& & $\sigma_{\theta}$ [mrad] & $\sigma_{\phi}$ [mrad] & $\sigma_{E_e}$ [keV] \\
\hline  
\hline  
Beryllium & 85 & 2.6 & 2.8 & 20 \\
Polyethylene & 128 & 2.7 & 2.8 & 20 \\
Scintillator (PVT)  & 125 & 2.8 & 3.2 & 20
 \\
\hline
\end{tabular}
\end{center}
\end{table*}

%
\begin{table*}[h!t]
\begin{center}
\caption{\label{tab:straggling_ms_gas} Relative $\theta_{e\gamma}$ and $\phi_{e\gamma}$ angles and energy $E_e$ uncertainties introduced by tracking (first figure) and material effects
between the target and the tracking detector (second figure), under different scenarios.  In the second and third scenario, the gas in the tracking volume 
extends to the region around the target to avoid a separation wall. The first tracking layer is placed at $R_e = 20$~cm. The photon detector is placed at 
$R_\gamma = 30$~cm. The tracking resolutions of MEG-II are assumed. Notice that, due to the correlations among variables, tracking and material contributions do not 
decouple completely | increasing the material effects also increases the impact of the tracking resolutions.}
\begin{tabular}{lccc}
\multirow{2}{*}{\bf Conditions} &\multicolumn{3}{c}{\bf Resolutions}\\ 
 & $\sigma_{\theta}$ [mrad] & $\sigma_{\phi}$ [mrad] & $\sigma_{E_e}$ [keV] \\
\hline
\hline
Pure Helium + 25 $\mu$m Kapton wall & 6.0 $\oplus$ 3.3 & 4.5 $\oplus$ 3.3 & 100 \\
Helium:CO$_2$ (90:10) & 6.0 $\oplus$ 3.2 &  4.6 $\oplus$ 3.4 & 100 \\
Helium:C$_{4}$H$_{10}$ (85:15) & 6.6 $\oplus$ 4.7 & 5.5 $\oplus$ 3.9 & 100 \\
\hline
\end{tabular}
\end{center}
\end{table*}

With the photon conversion technique, the photon conversion point can be measured very precisely, essentially with the single hit resolution of the 
$e^+e^-$ tracker ($\sim 100~\mu$m for a gaseous detector), with a first layer placed just behind the converter. 
As a consequence, the photon angle resolution
is completely dominated by the positron vertex resolution. With calorimetry, the detector and readout granularity of the 
entrance surface determines the resolution, but we can generally say that a sub-mm level can be reasonably reached, giving $\sim 1$~mrad 
contributions to $\phi_{e\gamma}$ and $\theta_{e\gamma}$ when the detector is placed at a few dozens of cm from the target (60 cm in MEG and MEG-II).

It must be noticed that an absolute calibration of $\Theta_{e\gamma}$ is very difficult to be obtained, resulting in a systematic uncertainty of a few mrad. 
As an example, in the MEG configuration, the 500~$\mu$m accuracy obtained for the target position along its normal direction translates into an uncertainty 
$> 3.5$~mrad on $\phi_{e\gamma}$.  Hence, quoting  angular resolutions at the mrad level is subject to the non-trivial ability of aligning the target with $\sim 100~\mu$m accuracy.

\subsection{Relative time $T_{e\gamma}$ }
\label{sec:Teg}

A  $T_{e\gamma}$ resolution of 120 ps  was  obtained in MEG with scintillation detectors, and  a 80 ps resolution is expected for MEG-II. 

The  positron time resolution of $\sim 35$~ps foreseen for MEG-II  might be probably improved with the incremental progress of the technologies. 
It is important to stress here that the positron time is usually measured at the end of the spectrometer, and the time of flight from the target to the timing detector 
needs to be subtracted. If there are long segments of the positron path which are untracked, the extrapolated track length can fluctuate significantly due to MS 
and $\Delta E$ in the crossed materials. In MEG, the track length uncertainty ($\sim 90$~ps) turned out to be the largest contribution to the $T_{e\gamma}$ 
resolution, due to the long untracked path ($\sim 1$~m) from the last reconstructed hit to the positron timing detector. In MEG-II this issue will be solved, thanks
to the 2~m-long drift chamber, which largely reduces  the untracked segments of the positron trajectory with respect to MEG.
Future designs should keep in mind this lesson, and this aspect could be critical for silicon detectors, which would track only a small portion of the 
positron trajectory.

For photons, if the conversion technique is adopted, a further complication arises. As far as only one conversion layer is foreseen, one can place thick 
scintillators at some distance from the converter, in such a way that either the electron or the positron reaches this detector. On the other hand, 
in order to stack multiple layers, a layer of active material just behind the converter should provide the required timing resolution.
A thick layer 
(few mm) of plastic scintillators would deteriorate unacceptably the  $E_\gamma$ resolution, and thin scintillating fibers (few 100~$\mu$m) cannot provide a resolution below a few 
100~ps and efficiencies above 90\%~\cite{atar}. Hence, a technological breakthrough is needed here, and a novel idea will be proposed in Sec.~\ref{sec:gamma_future}. For calorimetry, 
performances comparable or better than the MEG-II ones could be easily reached, considering the light  yield and  decay time of the  state-of-the-art scintillators.

\subsection{Summary}

Tab.~\ref{tab:limits} shows a summary of the limiting factors for the efficiency and resolutions of future \meg~searches. We stress again that, with the only exception of the 
tracking resolutions, these factors come from experimental conditions which are quite independent of the specific detector design, and only marginally dependent on the 
intrinsic resolutions of the detectors. It means that, even if the detector performances could be arbitrarily improved, most of these factors would remain unchanged. 
Hence, a radically new experimental approach would be needed to bring the resolutions on the \meg~discriminating variables significantly below these limits.

\begin{table*}[h!]
\begin{center}
\caption{\label{tab:limits} Limiting factors for the efficiency and resolutions of future \meg~searches.}
\begin{tabular}{lccc}
 & \multicolumn{2}{c}{\bf Typical figure} & \multirow{2}{*}{\bf Comments} \\
 & \emph{Calorimetry} & \emph{$\gamma$ Conversion} & \\
\cline{2-3}
\multicolumn{4}{l}{\emph{Efficiency}}\\
\hline  
\hline  
Material budget & 0.5 $\sim$ 0.9 & -- & magnet coil \\
Pair production & -- & 0.02 $\sim$ 0.04 & 0.05 $\sim$ 0.1 $X_0$\\
Minimum $e^+e^-$ energies &  -- & 0.8  & $E_{e^+}, E_{e^-} > 5$~MeV \\
 & & & \\
\multicolumn{4}{l}{\emph{Photon Energy Resolution}}\\
\hline 
\hline 
Energy loss & -- & 250 $\sim$ 800~keV & 0.05 $\sim$ 0.1 $X_0$\\ 
Photon Statistics \& segmentation & 800~keV & -- & \\
 & & & \\
\multicolumn{4}{l}{\emph{Positron Energy Resolution}}\\
\hline 
\hline 
Energy loss & \multicolumn{2}{c}{15~keV} & \\  
Tracking \& MS & \multicolumn{2}{c}{100~keV} & \\ 
 & & & \\
\multicolumn{4}{l}{\emph{Relative Angle Resolution}} \\
\hline 
\hline
MS on target & \multicolumn{2}{c}{2.6~/~2.8~mrad ($\theta_{e\gamma}$ / $\phi_{e\gamma}$)} & \\ 
MS on gas \& walls & \multicolumn{2}{c}{3.3~/~3.3~mrad ($\theta_{e\gamma}$ / $\phi_{e\gamma}$)} & \multirow{2}{*}{$R_e = 20$~cm, $R_\gamma = 30~$cm, B = 1~T} \\
Tracking & \multicolumn{2}{c}{6.0~/~4.5~mrad ($\theta_{e\gamma}$ / $\phi_{e\gamma}$)} & \\
Alignment & \multicolumn{2}{c}{$< 1$ mrad} & $<$100~$\mu$m target alignment\\
 & & & \\
\hline
\end{tabular}
\end{center}
\end{table*}

\section{Photon reconstruction perspectives}
\label{sec:gamma_future}

In this Section we discuss two possible realistic  photon detectors for \meg~searches. We will
consider a calorimetric approach with LaBr$_3$(Ce) crystals and a pair production approach with one or more layers
of conversion material.

\subsection{Calorimetry}
\label{sec:calo}

A homogeneous scintillation detector is placed out of the positron tracking volume and the magnetic field, and provides the $E_{\gamma}$, the photon conversion point, and the photon time measurements. With their high light yield and fast response, LaBr$_3$(Ce) crystals are a good candidates. Thanks to its high density (5.08~g/cm$^3$), a 20~cm long crystal with 13~cm diameter would contain the electromagnetic shower up to 100~MeV.
Silicon photon detectors like MPPCs could be coupled to the crystal, in such a way that a good coverage is guaranteed  ($\sim 50\%$ of  the crystal surface
considering the inactive areas of the single detector). 

We performed a set of GEANT4 simulations, which were validated against data obtained with  a 3 inch (diameter) $\times$ 3 inch (length) LaBr$_3$(Ce) crystal, 
irradiated with different sources, and in particular 9~MeV $\gamma$ rays from neutrons captured on Nickel, and instrumented with PMTs and MPPCs in order to characterize both 
the crystal and photon detector response ~\cite{papa}. The simulation includes the MPPC response, a full electronics chain and the reconstruction algorithms. Different geometries, 
sensors and analysis algorithms have been investigated. In the end, a  $\sigma_{\gamma}/E_{\gamma} \sim 1.6\%$ and a time resolution $\sigma_t \sim 30$~ps are predicted at the 
\meg~decay energy. Tab.~\ref{tab:baseline_calo} summarizes the performance of this solution.

\begin{table*}[h!]
\begin{center}
\caption{\label{tab:baseline_calo} Photon reconstruction performances of a baseline \meg~experiment with calorimetry.}
\begin{tabular}{lcc}
 \multicolumn{2}{c}{\bf Performance} & {\bf Source} \\
\hline
\hline
Acceptance & 70\% & \\
Efficiency & 60\% & \\ 
Photon Energy Resolution & $800~\mathrm{keV}$ & Energy loss \\
Photon Angle Resolution & 4.5 / 2.7 mrad ($\theta_{\gamma}$ / $\phi_{\gamma}$) & Positron vertex resolution \\
Photon Time & 30~ps & \\
\hline
\end{tabular}
\end{center}
\end{table*}

\subsection{Pair production}
\label{sec:conversion}

A basic design for a \meg~experiment adopting the photon conversion technique would consist of tracking detectors interleaved with one or more thin conversion layers. 
The design of the detector is also constrained by the requirement that an extended tracker for 52.8~MeV positron and tracking devices for the low-momentum 
photon-conversion products should coexist in the same magnetic field.

According to our results (see Sec.~\ref{sec:Eg}), the $E_\gamma$ resolution is expected to be dominated by the fluctuations of 
the energy loss in the converter, when its thickness is greater than 0.1~$X_0$. 
For smaller values, the tracking of the $e^+e^-$ pair can be relevant,
considering that previous studies~\cite{caltech,snowmass} point toward a~200-300~keV contribution.

Moreover, it should be noticed that, if a single layer is used, its size and the magnetic field can be optimized in such a way that the positron from the muon decay and at least 
one of the tracks in the $e^+e^-$ pair reach the outer radius of the detector, where fast detectors for timing could be placed.  If multiple layers are foreseen, the conversion layer itself should include an active component, able to measure the $e^+e^-$ timing with the required 
resolution (but scintillating fibers cannot provide the required performances).

A possible solution is given by a new generation of silicon detectors, with extremely good  time resolution. An R\&D activity  is ongoing  
(TT-PET project~\cite{TT-PET,TT-PET-test}) to realize a thin monolithic detector (100-300 um) in a Si-Ge Bi-CMOS process, that contains both the silicon 
sensor and the front-end electronics, featuring less than 100 ps time resolution for minimum ionizing particles. A dedicated design
could be adopted in the \meg~application, by stacking multiple detector layers in order to improve the resolution accordingly.

The additional low-Z material of the detector behind the converter is expected to deteriorate the $E_\gamma$ resolution without contributing significantly to the conversion
efficiency. According to our simulations, a single layer with the specifications in~\cite{TT-PET-test} (100~$\mu$m silicon on top of 50~$\mu$m Kapton) would
give a negligible contribution to the $E_\gamma$ resolution. For a 4-layer system, which would give a 50~ps resolution, comparable with the timing performances of the MEG-II
LXe calorimeter, this contribution would be $\sim 300$~keV, i.e. of the same order of the energy loss fluctuations for a 0.05~$X_0$ Lead converter.

It is also worth mentioning that, if the converter layer itself would be active and could provide some information about the energy deposit, it could be used 
to improve the $E_\gamma$ resolution.

\begin{table*}[h!]
\begin{center}
\caption{\label{tab:baseline_conv} Photon reconstruction performances of a baseline \meg~experiment with photon conversion.}
\begin{tabular}{lcc}
 \multicolumn{2}{c}{\bf Performance} & {\bf Source} \\
\hline
\hline
Acceptance & 70\%  & \\
Conversion Efficiency & 2.2\% & \\
$E_{e^+},E_{e^-} > 5$~MeV Selection Efficiency & 80\% & \\ 
Photon Energy Resolution & $250~\mathrm{keV} \oplus 200~\mathrm{keV}$ & Energy loss $\oplus$ tracking\\
Photon Angle Resolution & 4.5 / 2.7 mrad ($\theta_{\gamma}$ / $\phi_{\gamma}$) & Positron vertex resolution \\
Photon Time & 50~ps & \\
\hline
\end{tabular}
\end{center}
\end{table*}

Based on the results of Sec.~\ref{sec:factors}, Tab.~\ref{tab:baseline_conv} shows the expected photon reconstruction performances for this design. 
We assume:
\begin{itemize}
\item one passive conversion layer, 0.05~$X_0$ Lead, covering $135^\circ$ in $\phi$ and 60~cm in $Z$, and placed at $R_\gamma = 30$~cm;
\item scintillating tiles at the end of the trajectory of conversion pairs
and positrons from muons, providing a time resolution of 
$\sim$~50~ps;
\item a tracking system providing an $e^+e^-$ vertex resolution which contributes negligibly to the photon angle uncertainty (with respect to the contribution
of the positron vertex reconstruction) and to the  $E_\gamma$  resolution (with respect to the fluctuations of the energy loss in the converter).
\end{itemize}
If multiple conversion layers are present and timing is provided by the TT-PET detectors, the 50~ps resolution can be preserved but the photon energy resolution deteriorates as estimated above.

\begin{figure}
\resizebox{0.5\textwidth}{!}{
  \includegraphics{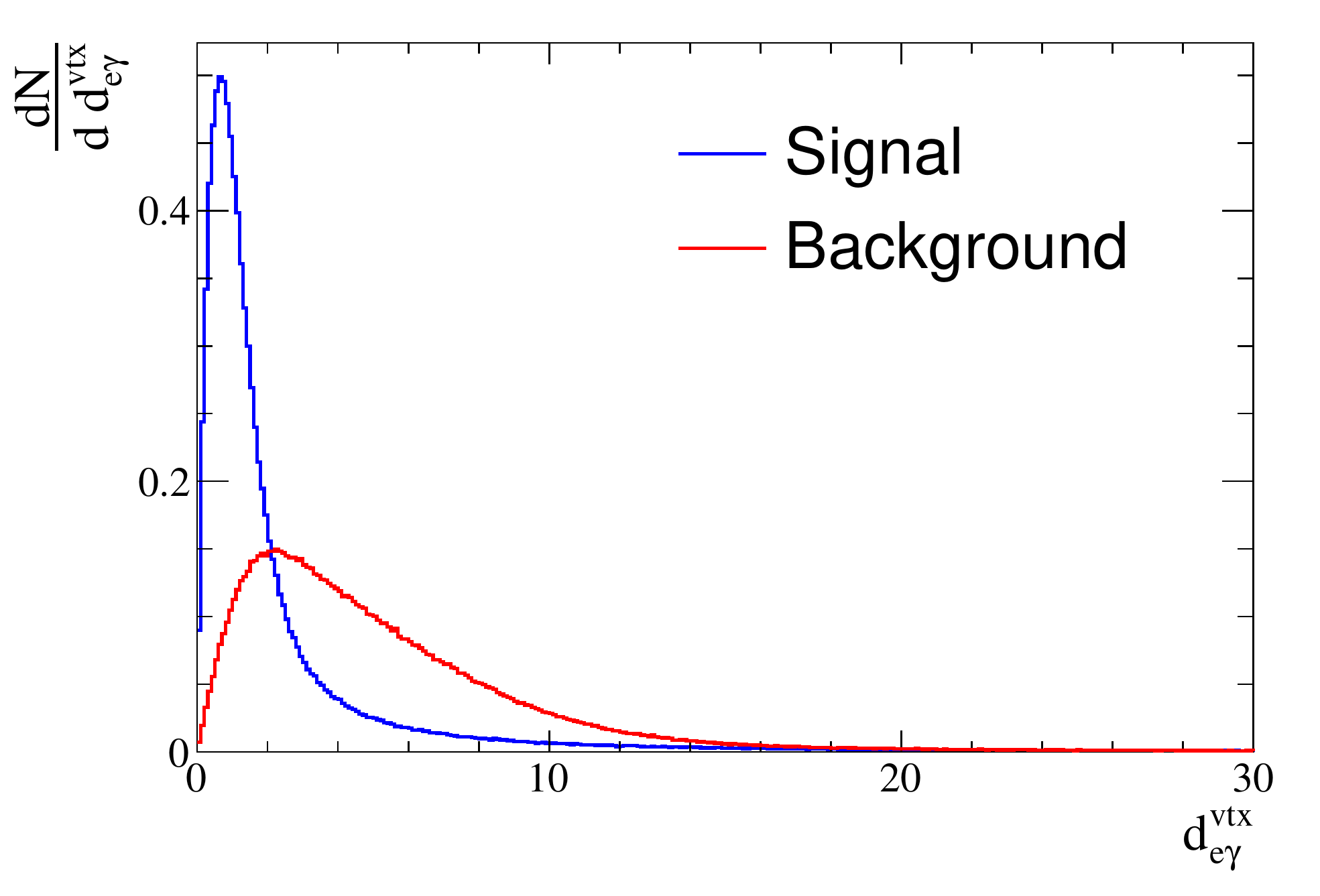}
}
\caption{Distribution of the normalized distance between the positron and photon vertices, for signal (blue) and accidental background (black) events, with a $0.1 X_0$ Lead converter.}
\label{fig:vertexing}
\end{figure}

Beside providing better resolutions with respect to calorimetry, the photon conversion technique also provides a measurement of the photon 
direction, from the combination of the reconstructed directions of the $e^+e^-$ pair, independently of the positron reconstruction. The
resulting angular resolution, deteriorated by the MS in the converter, is $\sim~80$~mrad with $0.1 X_0$ Lead, and hence cannot compete with the 
one obtained from the combination of the positron vertex and photon conversion point. Nonetheless, this additional information can be used to reduce 
the accidental background. In Fig.~\ref{fig:vertexing} we show the distribution of the normalized distance defined as:
\begin{equation} 
d^{\mathrm{vtx}}_{e\gamma} = \sqrt{\left(\frac{X_e - X_\gamma}{\sigma_X} \right)^2 + \left(\frac{Y_e - Y_\gamma}{\sigma_Y} \right)^2}
\end{equation} 
where  ($X_e$ , $Y_e $) and ($X_\gamma$ , $Y_\gamma $)  are the coordinates on the target of the vertex obtained in two ways, one by propagating the positron track back to the target 
and the other by using  the direction of the  $e^+e^-$ pair. The uncertainties ($\sigma_X$ and  $\sigma_Y$) of the  $e^+e^-$ back-propagation are used, since they largely dominate over the positron vertex resolutions. The expected distributions for signal and accidental background events with the photon coming from the target (assuming 
the same beam profile used in MEG) are shown in Fig. ~\ref{fig:vertexing}, for a 0.05~$X_0$ Lead converted at $R_\gamma = 30$~cm, a $76^\circ$ slanted target and the acceptance defined in Sec.~\ref{sec:Thetaeg}. 
In this scenario, the optimal ratio of signal to square root of background is obtained for $d^{vtx}_{e\gamma} < 1.25$,
which removes 91\% of background events with 52\% signal efficiency. Similarly, background produced by positron AIF occurring far from the target would be easily removed 
without any significant loss of signal efficiency. However the average background rejection capability is lower if multiple conversion layers are used, because the 
layers after the first one have a larger $R_\gamma$ and hence the resolution of the back-projection to the target is worse for photons converting there.

\section{Positron reconstruction perspectives}
\label{sec:e_future}

We consider  a positron detector, composed of two sectors: a vertex detector for a precise determination of 
the muon decay point and the positron angles, and an extended tracker for the measurement of the positron momentum.

\subsection{Vertex detector}

As already discussed, a silicon detector would not be competitive with a gaseous detector as an extended tracker. Nonetheless, 
we can still consider the possibility of having two layers of silicon detectors for vertexing. Silicon pixels would give a very good
vertex resolution, thanks to the very precise determination of both the azimuthal and the longitudinal coordinate ($\sim~10~\mu$m). In practice,
the vertex and angle resolution would be completely dominated by the MS in this detector. As a consequence, the extended tracker  would be only useful for the determination of $E_e$.

As an alternative, one could consider a time projection chamber (TPC) with a very light (helium-based) gas mixture. The single hit resolution of such
a device would be limited by the diffusion of the drifting electrons, but a large number of hits would be available.
Gaseous electron multiplier (GEM) foils or Micromegas could be used to generate the electron avalanche, inducing signals on readout pads and allowing  the TPC to be 
operated in continuous mode (no gating) even in presence of a very high track rate~\cite{gem_tpc}.

 We performed simulations with the GARFIELD 
software~\cite{garfield}, assuming a He:CO$_2$ (90:10) gas mixture, a 0.5~T magnetic field and a 1~kV/cm electric field. We assume the readout to be 
performed with very high granularity, as in the GEMPIX~\cite{gempix} and InGrid~\cite{ingrid} projects, so that the single ionization clusters are detected 
and the space resolution is dominated by the diffusion of the drifting electrons. For a given drift distance $d_{\mathrm{drift}}$, a resolution of 
$\sim 180 (150) ~\mu\mathrm{m} \cdot \sqrt{d_{\mathrm{drift}} [cm]}$ is found in the azimuthal (longitudinal) coordinate, with $\sim 100$ hits per track. 
On the other hand, at a very high rate, the use of such a device would be limited by both the rate capability of the multiplication stage and the space 
charge accumulated in the drift region due to the primary ionization itself. 

\subsection{Extended tracker}

The basic option for the extended tracker is a drift chamber, with stereo wires for the measurement of the longitudinal position. The MEG-II
drift chamber can be used as a benchmark for the material budget and the single hit resolution.

On the other hand, the high track rate in the inner layers is expected to produce visible aging effects at the beam rate expected in 
MEG-II~\cite{meg2}. It could make a TPC the only choice for a gaseous extended tracker at higher beam rates. The detector geometry would be strongly 
constrained, because a very long TPC ($\sim$~1~m drift distance) could not provide acceptable resolutions due to the electron diffusion. 
As an alternative, in the early stages of the MEG upgrade project, a 2~m long radial TPC was proposed, with a He:CO$_2$:C$_2$H$_6$ (70:10:20) 
gas mixture. The radial design implies some technical difficulties connected to the drift of the electrons orthogonally to the magnetic field. First,
their diffusion is not suppressed by the magnetic field as in a longitudinal TPC. Second, the curved trajectory of the drifting positrons need to be 
accounted in the reconstruction stages. If these problems can be overcome with a proper tuning of the gas mixture composition, an accurate knowledge of 
the magnetic field and a detailed calibration of the drift trajectories, a resolution of $\sim 130~\mu\mathrm{m} \cdot \sqrt{d_{\mathrm{drift}} [cm]}$ in the radial 
and azimuthal coordinates could be achieved~\cite{meg2}, with drift distances not exceeding 10~cm.

\section{Target optimization}
\label{sec:target}

The target thickness represents one of the most stringent limitations to the achievable angular resolution. In order to use thinner targets, two options can 
be investigated.
\begin{enumerate}
\item If the muon momentum can be significantly reduced, without reducing the beam intensity and preserving a good momentum bite, the distribution of the muon decay
depth in the target (i.e. the width of the Bragg peak) is reduced accordingly. A thinner target could be used, improving the angular resolution without affecting  $\Gamma_\mu$, $\epsilon_e$ and $\epsilon_\gamma$.
\item The target could be replaced by multiple thinner targets. A tentative design consists of a V-shaped target, made of two planes forming a $152^\circ$ angle. The problem with
such an approach is that a relevant fraction of muons would decay in the gas within the two targets: signal events from such muons could not be identified, contributing only to the accidental background. As an example, two Beryllium targets of 40~$\mu$m thickness each would provide 
80\% stopping efficiency on target, while 13\% of muons would decay in the gas between the two target sections. The advantage in terms of angular resolutions would be far too small to compensate this inefficiency. The use of multiple targets can nonetheless help to reduce the background, when the photon conversion approach is 
used to reconstruct the photon direction. If, for instance, the single target foil is replaced by two staggered foils, each 
illuminated by half the beam spot, and with sufficient separation in space, the back-propagation of the $e^+e^-$ pair can be used to identify the
foil where the photon has been produced, and check if it is the same of the positron. 
In this case, the accidental background is effectively reduced by a factor of two. More generally, 
spreading the beam over a larger surface makes more effective the background rejection based on the goodness of
the electron-photon vertex.
\end{enumerate}

\section{Sensitivity reach}
\label{sec:sens}

In this Section we give an estimate of the sensitivity reach of a \meg~search based on the technologies described above. At first, we
consider a basic design  based on the photon conversion technique, with a single conversion layer, an inner vertex detector (silicon pixels or a TPC) and a 200~cm long extended
tracker (a drift chamber or a TPC) which would serve as a positron and a positron-electron pair spectrometer. 

The inclusion of multiple conversion layers would be an interesting improvement to this design. It can be made without any loss in the timing performances 
only if timing is provided by fast silicon detectors at the conversion layer. 

We finally consider a calorimetric approach for the photon reconstruction, while leaving the positron reconstruction unchanged.

In both cases, we neglect the difficulties connected to the reconstruction of signal events in a crowded environment  with positron tracks from multiple Michel muon decays.

\subsection{A  design with photon conversion}

\begin{figure*}
\centering
\resizebox{\textwidth}{!}{
  \includegraphics{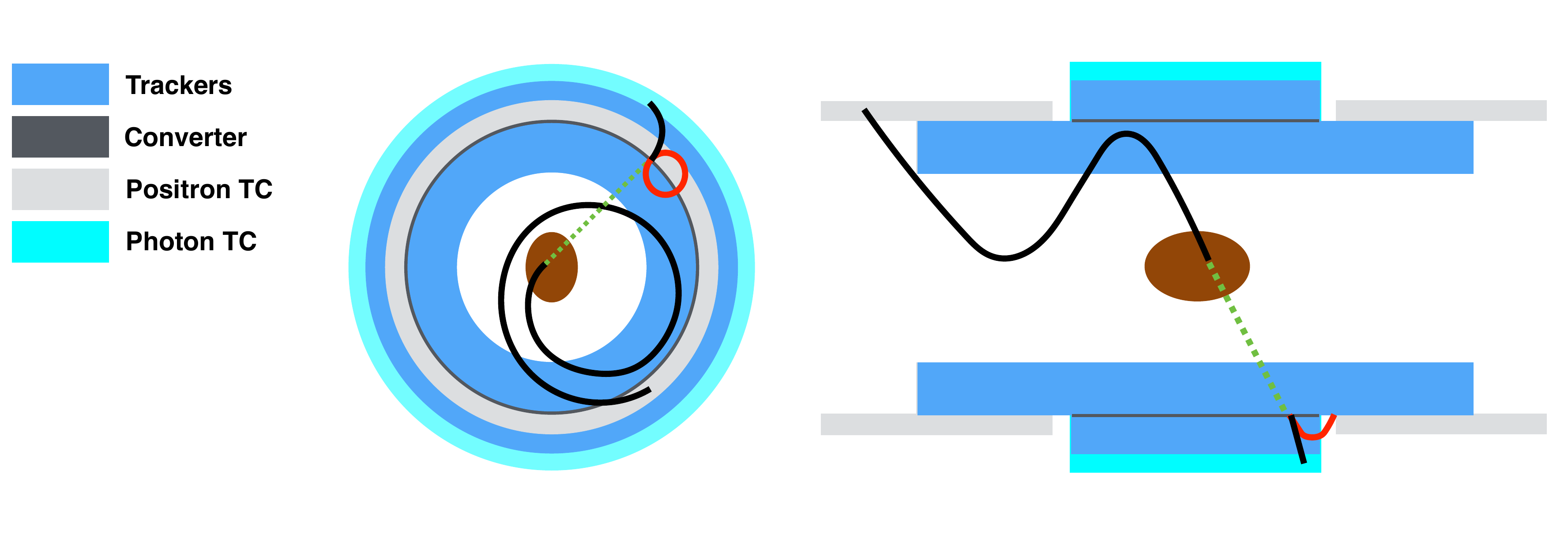}
}
\caption{Sketches of a detector design made of an extended $e^+$ and an $e^+ e^-$ pair trackers (sky blue) separated by a thin conversion layer (dark gray, not in scale),
with positron (light gray) and photon (cyan) timing counters (TC). A typical $\meg$~event with converted photon is shown (positrons in black, photon in green, electron in red).}
\label{fig:sketch_sens}
\end{figure*}

In Fig.~\ref{fig:sketch_sens} we show a sketch of a \meg~detector based on the photon conversion technique, with two different options 
for the inner vertex detector and a typical signal event. A similar design was recently proposed in~\cite{snowmass}.

In this design, a target identical to the one of MEG-II is surrounded by a positron tracker extending from $R = 20$ to $R = 30$~cm.
with a length of 200~cm. It can be a drift chamber or a radial TPC. As in MEG and MEG-II, plastic scintillators (positron timing counters) 
are placed behind it, in order to measure the positron track timing.

At $R = 30$~cm, 
a 60~cm long Lead conversion layer is placed, with a $0.1~X_0$ thickness. The 
longitudinal extent of the conversion layer defines the acceptance of the detector, $\sim 70\%$.

Externally, a 84~cm long drift chamber or radial TPC is used as an electron-positron pair spectrometer. This chamber extends up to $R = 42$~cm, where 
plastic scintillators (photon timing counters) are placed.

Optionally, a small TPC or a two-layer silicon vertex detector can be considered. Both detectors are 40~cm long. 
The TPC has an inner radius of 10~cm and an outer radius of 20~cm. 
The first silicon layer is placed at a radius of 10~cm. 

Everything is immersed in a graded magnetic field similar to the MEG one, such that, for events within the acceptance defined above, the signal positron curls before reaching the
converter layer and finally reaches the positron timing counters, while at least one of the tracks from the photon conversion goes through the  whole  $e^+e^-$pair spectrometer 
and reaches the photon timing counters.

We estimated the expected performances of such a detector. For simplicity, we rely on the results shown in Tab.~\ref{tab:baseline_conv} for the photon reconstruction,
although they are obtained for a uniform magnetic field. For the positron  angle and  momentum reconstruction in the tracker we assume the performances 
of the MEG-II drift chamber, with a 90\% reconstruction efficiency, while for the vertex resolution with an inner tracker
we assume two different scenarios. In the first, conservative one, the only improvement comes from having the first measured point which is closer to the target, 
while the momentum and angular resolutions are still dominated by the extended tracker, and the angular resolution is deteriorated by the presence of the inner wall 
of the TPC or the inner layer of the silicon vertex tracker. In the second, optimistic one, the vertex detector makes
the tracking contribution to the angular resolution negligible. This resolution is then completely determined by material effects before and inside the first 
layer of the inner vertex detector. A summary of the expected performances can be found in Tab.~\ref{tab:perf_ee_1}  and ~\ref{tab:perf_ee_2}. 
It is evident that a silicon vertex detector cannot help, because the MS in the first layer of such a detector negates the advantage of having a very good 
determination of the track angle between the first and the subsequent layers.

\begin{table}[h!t]
\begin{center}
\caption{\label{tab:perf_ee_1}  Expected performances (efficiency and resolutions)  for a basic design with different options as discussed in the text.}
\begin{tabular}{lcc}
{\bf Observable} &  \emph{one photon } &  \emph{ photon }  \\
                          &  \emph{conversion layer} &  \emph{  calorimeter}  \\
\hline
$T_{e\gamma}$ [ps] & 60  & 50 \\
$E_e$ [keV] & 100   & 100 \\
$E_\gamma$ [keV] & 320   & 850 \\
Efficiency [\%] & 1.2   & 42 \\
\hline
\end{tabular}
\end{center}
\end{table}

\begin{table}[h!t]
\begin{center}
\caption{\label{tab:perf_ee_2}  Angular resolutions for different types of a vertex detector. A conservative estimate is given in parenthesis.}
\begin{tabular}{lcc}
                          &  $\theta_{e\gamma}$  [mrad] &  $\phi_{e\gamma}$  [mrad] \\
\hline
None   & 7.3  & 6.2   \\
  TPC  &  3.5 (6.1)  & 3.8 (4.8)   \\
   Silicon  &  8.0 (6.3) & 7.4 (6.9) \\
\hline
\end{tabular}
\end{center}
\end{table}

We also considered a simpler design, with similar radial dimensions, where the magnetic field is reduced to 0.5~T and the conversion layer covers only a portion ($\sim 18\%$)
of the azimuthal angular range. In this design, signal positrons reach the  $e^+ e^-$pair spectrometer, which also acts as an extended tracker, without hitting the conversion layer
if the corresponding photon does. Finally, the positron reaches the same counters used for photon timing. Simulations show that such a design, similar to the
one proposed in~\cite{caltech}, implies a large degradation of the momentum resolution. While it could be still suitable for a beam rate $\sim 10^9~\mu$/s, 
this design would not fit a larger rate and, as acknowledged in~\cite{caltech}, the optimal working point would be even lower in a scenario where multiple
conversion layers are used.

\subsection{A design with calorimetry}

A \meg~experiment based on calorimetry could have a design very similar to the one above for the central part of the detector, but the external  $e^+ e^-$ pair tracker would be replaced
by a scintillation detector placed outside of the magnet. With LaBr$_3$(Ce) crystals, the calorimeter could be about 20~cm deep and the performance summarized in 
Tab.~\ref{tab:perf_ee_1}  and ~\ref{tab:perf_ee_2}  could be reached. Here we assume that the photon conversion point can be still determined with a negligible resolution compared to the 
positron vertex resolution.

\subsection{Sensitivity estimates}

We consider here 100 weeks of data taking (3 to 4 years at PSI), with muon rates from $10^{8}$ to $10^{10}$ muons per second. We define a different signal region for
each scenario, in such a way that, according to the resolutions estimated above, the efficiency for the signal to be inside that region is always 70\%.
We then assume that a counting analysis is performed on the events falling within this region.

Formulas in~\cite{kuno-okada} allow to estimate the background rate, by using as an input the measured photon rate in the MEG calorimeter, linearly scaled with the beam rate. 
Considering that the geometry in the central region of the detector is very similar to the MEG one, this approach takes into account reliably the rate of AIF photons, which 
would be otherwise very difficult to extract from simulations, given the extremely low probability of this process per single muon decay.

When the photon conversion technique is adopted, we also assume a background rejection performed by requiring a 
good electron-photon vertex as explained in Sec.~\ref{sec:gamma_future}. The efficiency and background rejection capabilities of this approach 
are determined in each scenario according to the expected resolutions. 

Finally, we extract the expected sensitivity of the experiment according to a frequentistic approach~\cite{feldman-cousins}. 

\begin{figure*}
\centering
\resizebox{0.65\textwidth}{!}{
  \includegraphics{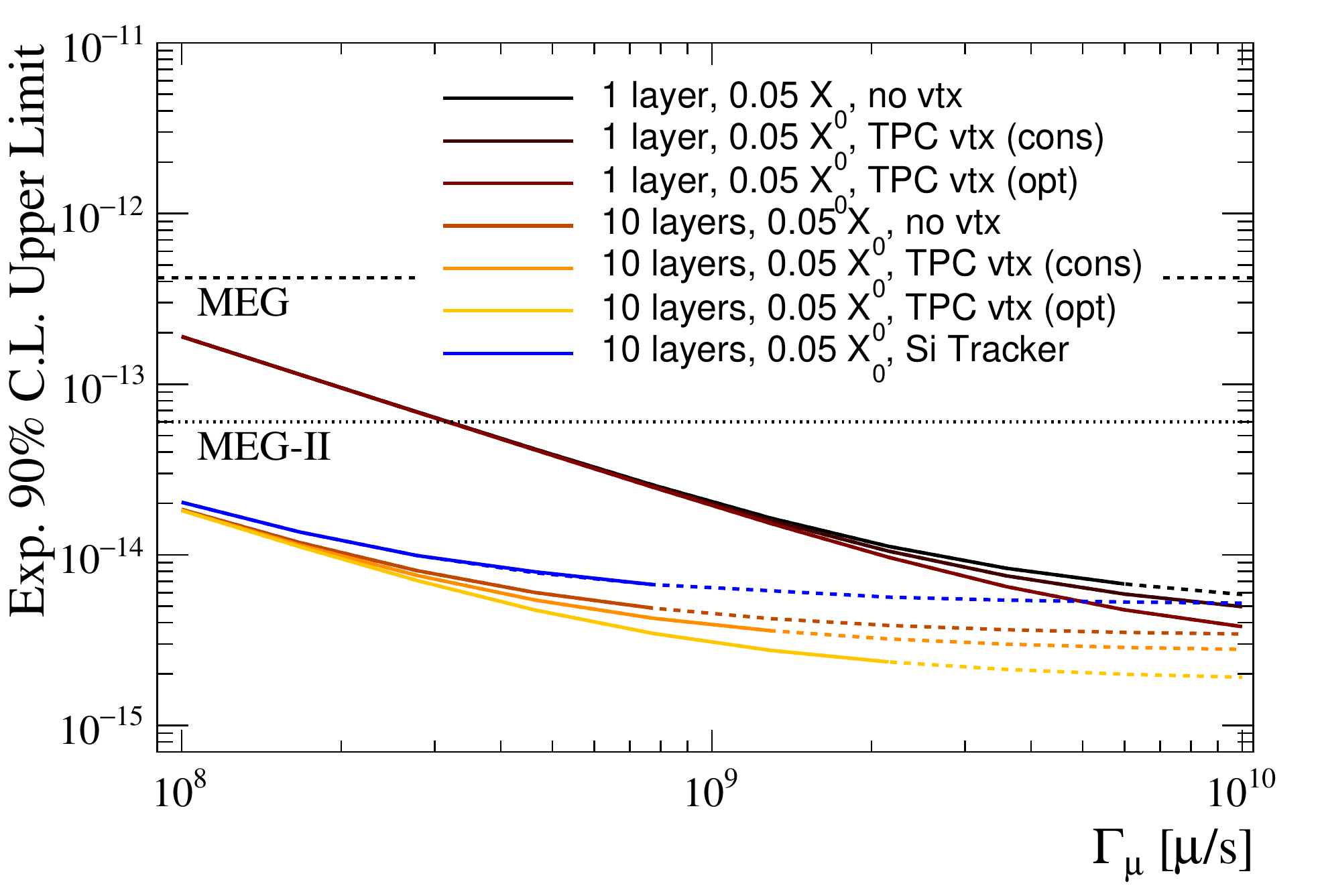}
}
\caption{Expected 90\% C.L. upper limit on the Branching Ratio of \meg~in different scenarios for a 3-year run. A few different designs based on the photon conversion technique 
are compared, including the TPC vertex detector option in the conservative and optimistic hypotheses. The lines turn from continuous to dashed when the number of background 
events exceeds 10. The horizontal dashed and dotted lines show the current MEG limit and the expected MEG-II sensitivity, respectively.}
\label{fig:sens1}
\end{figure*}

\begin{figure*}
\centering
\resizebox{0.65\textwidth}{!}{
  \includegraphics{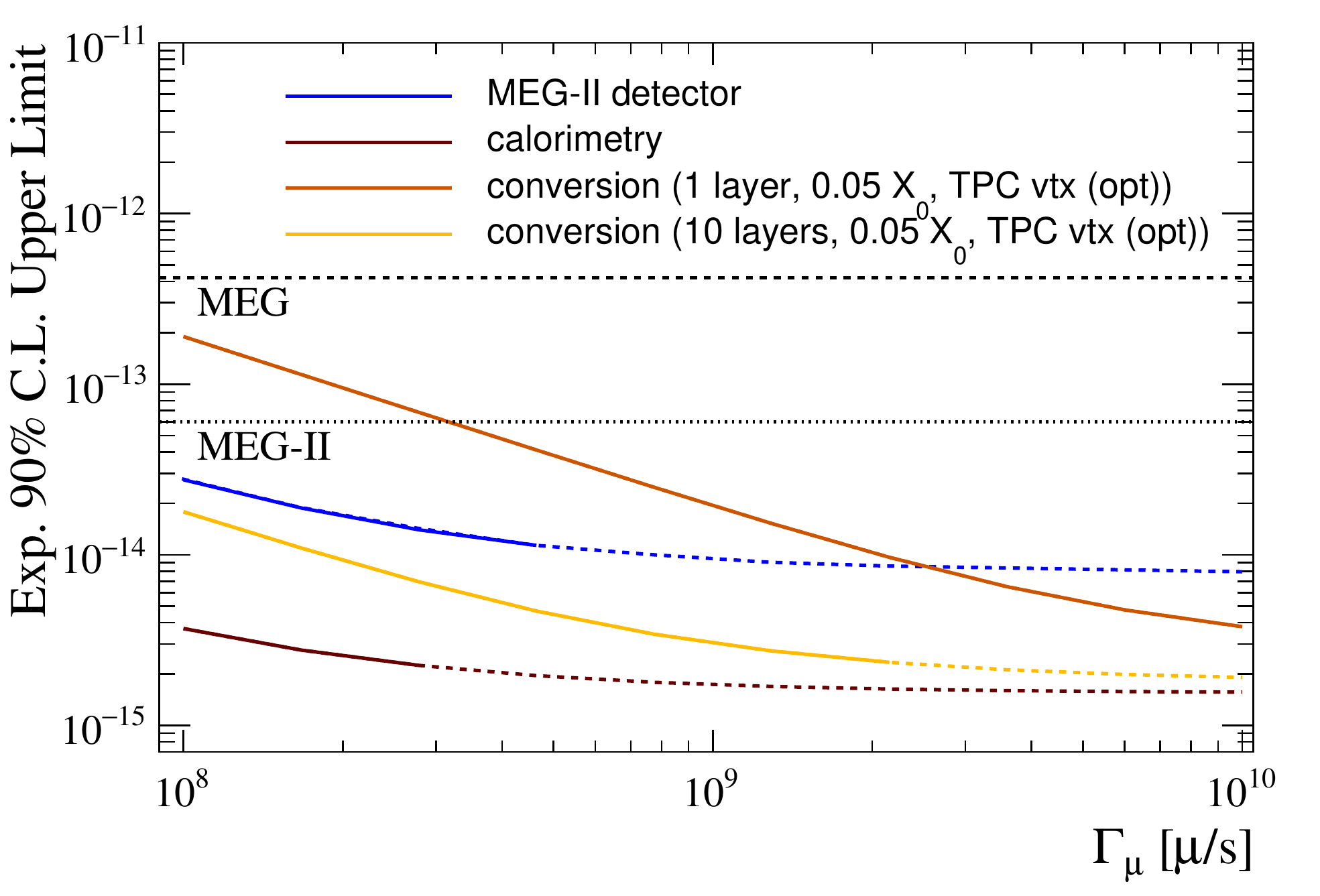}
}
\caption{Expected 90\% C.L. upper limit on the Branching Ratio of \meg~in different scenarios for a 3-year run. Calorimetery and the photon conversion technique are compared. 
The lines turn from continuous to dashed when the number of background events exceeds 10. The horizontal dashed and dotted lines show the current MEG limit and the 
expected MEG-II sensitivity.}
\label{fig:sens2}
\end{figure*}

Figures~\ref{fig:sens1} and \ref{fig:sens2} show the expected sensitivity to the \meg~decay as a function of the beam intensity in different scenarios. 

Among them, we also considered the possibility of multiple conversion layers. In this case, we introduce fast silicon detectors
for timing, with several layers  to reach a 50~ps time resolution. The
photon energy resolution is degraded accordingly (see Sec.~\ref{sec:gamma_future}).

In Fig. \ref{fig:sens1} we compare different designs based on the photon conversion approach. Apart from the obvious advantage of having multiple layers, it should be noticed that a 
vertex detector would be only useful at very large beam rates. We do not consider the silicon vertex detector option because, according to Tab.~\ref{tab:perf_ee_2}, it would not
significantly improve the expected performances. We consider instead a scenario where the extended tracker is made of silicon detectors, with the performances
presented in~\cite{caltech}, which could be the only available solution if aging effects make impossible to operate a gaseous detector.

In  Fig. \ref{fig:sens2} we compare the performances of an experiment with calorimetry with the performances  of 
the best photon conversion designs. We also show for comparison how
the MEG-II detector would perform at the same beam rates. 
Calorimetry is definitively advantageous at low beam rate, 
as expected, but there is a wide range of beam intensity where this approach would be limited by the background, while the photon conversion approach would not give yet 
a better sensitivity, unless a very large detector with many conversion layers is built. 

In conclusion, a $2 \times 10^{-15}$ limit seems to be within reach with a  $10^{9}$ muons per second stopping rate, while a further increase of the beam rate up
to $10^{10}$ would only improve the sensitivity by a factor of 2.

\section{Conclusions}

Efforts are ongoing to develop muon beam-lines with intensities near $10^9$ and possibly approaching $10^{10}$ muons per second, to be used for a future generation of 
cLFV searches in muon decays.The HiMB project at PSI aims to reach $10^{10}$ muons per second in the next decade, while the MuSIC project at RCNP (Japan)
is experimenting different approaches to increase the muon yield per unit of power of the primary proton beam. The FNAL project PIP-II 
could be also competitive in this field.

In this paper we investigated the experimental factors that will limit the sensitivity reach of future experiments
searching for the \meg~decay with a continuous  muon beam  at high intensity.

The most relevant issue is the choice of the photon detection technique between calorimetry  and the reconstruction
of the $e^+e^-$ pair from photon conversion in a thin layer of high-Z material, being favored  the former by the
much higher detection efficiency and the latter by the far superior resolutions, along with the
possibility of rejecting accidental background events by reconstructing the photon-positron vertex.

On the positron side, tracking with gaseous detectors would ideally provide the best possible resolutions, which would
be eventually limited by the multiple Coulomb scattering experienced by the particle in the target and in the material in front
of the tracker. On the other hand, the high occupancy in the inner part of the tracking system could severely limit the
possibility of using gaseous detectors. A significant deterioration of the overall sensitivity (more than a factor 2) is
expected if a silicon tracker has to be used for this reason.

Sensitivity projections show that a 3-year run with an accelerator delivering around $10^9$ muons per second could 
allow to reach a sensitivity of a few $10^{-15}$ (expected 90\% upper limit on the \meg~BR), with
poor perspectives of going below $10^{-15}$ even with $10^{10}$ muons per second. Below $5 \times 10^8$ muons per
second, the calorimetric approach needs to be used in order to reach this target. If a muon beam rate exceeding $10^9$
muons per second is available, the much cheaper photon conversion option would be recommended and
would provide similar sensitivities.

The sensitivity would be eventually limited by the fluctuation of the interaction of the particles with the detector materials: this indicates
that a further step forward in the search for \meg~would require a radical  rethinking of the experimental concept.

\section{Acknowledgments}
This paper is dedicated to the memory of our colleague Giancarlo Piredda, whose inspiration was invaluable in the early stages of this work.
We are also grateful to all our MEG and MEG-II colleagues for valuable discussions.

\end{document}